\documentclass[aps,prl,twocolumn,showpacs,amsfonts,superscriptaddress,floatfix]{revtex4}

\usepackage{amsmath}
\usepackage{graphicx}
\usepackage{multimedia}
\usepackage{color}
\usepackage{bbold }
\usepackage{bm}
\usepackage{float}

\newcommand{\beq}{\begin{equation}}
\newcommand{\eeq}{\end{equation}}
\newcommand{\beqarray}{\begin{eqnarray}}
\newcommand{\eeqarray}{\end{eqnarray}}

\begin{document}

\title{Persistent superfluid flow arising from the He-McKellar-Wilkens effect in molecular dipolar condensates} 
\author{A. A. Wood} 
\affiliation{School of Physics, University of Melbourne, Victoria 3010, Australia}
\author{B.H.J. McKellar} 
\affiliation{ARC Centre of Excellence for Particle Physics at the Terascale, School of Physics, University of Melbourne, Victoria 3010, Australia}
\author{A. M. Martin} 
\affiliation{School of Physics, University of Melbourne, Victoria 3010, Australia}
\date{\today}

\begin{abstract}
We show that the He-McKellar-Wilkens effect can induce a persistent flow in a Bose-Einstein condensate of polar molecules confined in a toroidal trap, with the dipolar interaction mediated via an electric dipole moment. For Bose-Einstein condensates of atoms with a magnetic dipole moment, we show that although it is theoretically possible to induce persistent flow via the Aharonov-Casher effect, the strength of electric field required is prohibitive. We also outline an experimental geometry tailored specifically for observing the He-McKellar-Wilkens effect in toroidally-trapped condensates. 
\end{abstract}


\maketitle
The experimental study of persistent superfluid flow in Bose-Einstein condensates (BECs) in toroidal traps \cite{PRL99_260401,PRL106_130401,PRA86_013629,PRL110_025301,PRL110_025302,PRA88_063633,PRL111_205301,PRL111_235301,PRL113_045305,PRL113_135302,NJP16_013046,Nature506_200} has matured significantly over the last decade.  As such, ring-shaped BECs in toroidal traps have been the subject of many experimental and theoretical investigations \cite{PRA66_053606,EuroLett46_275,JPhysB34_L113,SciRep2_352,PRA64_063602,PRA74_061601} focusing on persistent currents \cite{PRL99_260401,PRL110_025301,PRA88_051602}, weak links \cite{PRL106_130401,PRL110_025302}, formation of matter-wave patterns by rotating potentials \cite{PRA86_023832}, solitary waves \cite{JPhysB34_L113,PRA79_043602}, and the decay of the persistent current via phase slips \cite{PRA86_013629,PRA80_021601,JPhysB46_095302}. In these studies the persistent flow is created by transferring angular momentum from optical fields \cite{PRL99_260401,PRL110_025302} or by stirring with a rotating barrier \cite{PRL110_025302,PRA88_063633}. 

In this work we consider an alternative approach for the generation of a persistent flow for dipolar condensates in a toroidal trap. We show the He-McKellar-Wilkens effect \cite{PRA47_3424,PRL72_5} can induce a persistent flow in a BEC of molecules with a significant electric dipole moment confined in a toroidal geometry. We also find that for an atomic dipolar BEC, where the constituent atoms have a large magnetic dipole moment, the Aharonov-Casher effect \cite{PRL53_319, PRL48_1660,PRL63_380} could be used to generate a persistent flow. However, our calculations show that while it is feasible to use the He-McKellar-Wilkens effect to drive the creation of a persistent flow in a BEC of polar molecules, the electric field strengths required for the Aharonov-Casher effect in a magnetic dipolar BEC are prohibitive. 

\begin{figure}
\includegraphics[width=\columnwidth]{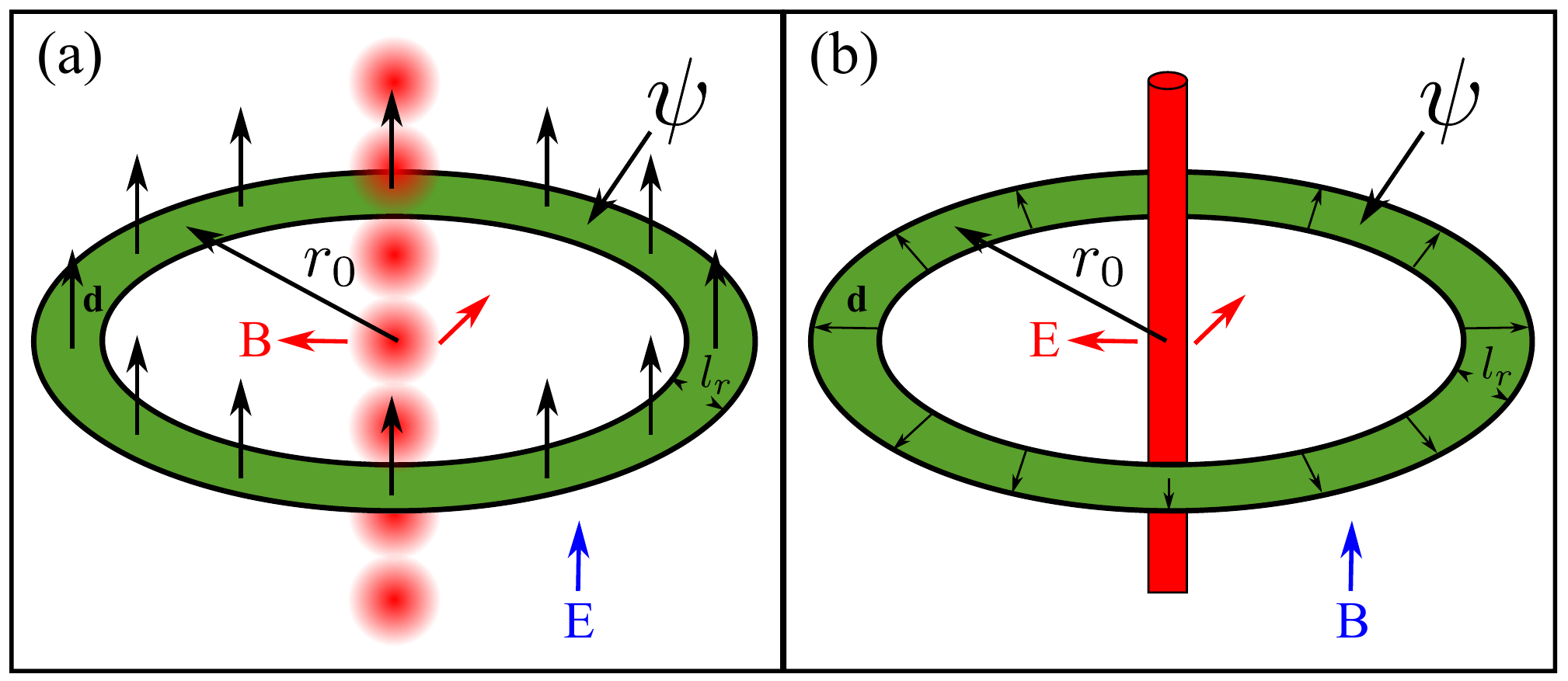}
   \vspace{-0.5cm}
   \caption{(Color online) (a,b) Schematic of a dipolar superfluid (green) in a toroidal geometry, with $r_0$ being the radius of the toroid and $l_r$ being the width of the superfluid in the toroid.  In (a) the electric dipole moment, ${\bf d}$, of the molecules in the superfluid (denoted by the arrows) is perpendicular to the plane (aligned via an external electric field) of the trap and reside in a magnetic field generated by an infinite line of magnetic monopoles (denoted by red spheres). In (b) the electric dipole moment, ${\bf d}$, of the molecules in the superfluid (denoted by the arrows) are aligned, via an infinitely long charged wire (red tube), radially (black arrows) and reside in a uniform magnetic field perpendicular to the plane.
   }\label{rydberg}
   \vspace{-0.5cm}
\end{figure}

The He-McKellar-Wilkens phase is the electromagnetic dual of the Aharonov-Casher \cite{PRL53_319, PRL48_1660,PRL63_380} geometric phase.
The Aharonov-Casher geometric phase arises when a magnetic dipole encircles an infinite line of electric charges. 
Its dual, as pointed out by He and McKellar in 1993 \cite{PRA47_3424}, arises when an electric dipole encircles an infinite line of magnetic monopoles, as shown in Fig. \ref{rydberg}(a).
 
The original work of He and McKellar did not suggest an experimental test for the observation of this geometrical phase due to the inherent difficulty in arranging a line of magnetic monopoles. 
However, an experimental proposal was later developed by Wilkens \cite{PRL72_5}, which considered the case of an electric dipole interacting with a magnetic field generated with ferromagnetic materials. 
Subsequently, Wei, Han and Wei \cite{PRL75_2071} showed that, for an induced electric dipole moment, it is possible to interchange the electric and magnetic fields \emph{i.e.} having a radial electric field generated by an infinite line of charge and a magnetic field perpendicular to the electric field and parallel to the line of charge.
The He-McKellar-Wilkens phase has recently been experimentally measured, by Lepoutre {\it et al.} \cite{PRL109_120404}, using an atom interferometer, for a field geometry equivalent to that proposed by Wei, Han and Wei \cite{PRL75_2071}. The same group later reported the experimental observation of the Aharonov-Casher phase \cite{EPJD_68_168}.

Bose-Einstein condensates have been formed from atoms with large magnetic dipole moments such as  $^{52}$Cr~\cite{PRL94_160401,PRA77_061601}, $^{164}$Dy~\cite{PRL107_190401}, and $^{168}$Er~\cite{PRL108_210401}. The effects of significant dipolar interactions observed in these gases, including magnetostriction \cite{Nature448_672}, d-wave collapse \cite{PRL101_080401} and the possible formation of a supersolid phase \cite{Nature530_194}. The formation and cooling to degeneracy of polar molecules, with significant electric dipole moments, is proving to be more challenging. However, recent significant progress has been made with $^{40}$K$^{87}$Rb~\cite{arXiv:0811.4618} and $^{133}\text{Cs}^{87}\text{Rb}$~\cite{PRL113_255301}.

In the original proposal for the He-McKellar-Wilkens \cite{PRA47_3424} geometric phase, an electric dipole moving on a path $\mathcal{C}$ that encircles a line of magnetic monopoles acquires a phase shift given by 
\begin{equation}
\phi_\text{HMW} = -\frac{1}{\hbar}\int_\mathcal{C} {\bf d}\times {\bf B} \cdot d{\bf l},
\label{eq:hmw_def}
\end{equation}
where ${\bf B}$ is the radial magnetic field from the line of monopoles.

One might expect that a BEC of electric dipoles in a toroidal geometry, with the electric dipoles aligned perpendicular to the plane of the toroid could exhibit persistent flow due to $\phi_\text{HMW}$, if a line of magnetic monopoles passed through the centre of the toroid, as schematically shown in Fig. 1(a). Magnetic monopoles have yet to be discovered, and as such this geometry is experimentally challenging. However, adopting the proposal from Wei, Han and Wei \cite{PRL75_2071}, a  persistent current, arising from the He-McKellar-Wilkens geometric phase can arise in the geometry schematically shown in Fig. 1(b). In this geometry the dipolar BEC is confined in a toroidal trap with the dipoles aligned radially, via an electric field generated from a line of charge passing through the toroid, and a uniform external magnetic field is aligned perpendicularly to the toroid. Such a geometry reveals an insightful link \cite{PRL75_2071} between the Aharonov-Bohm phase \cite{PR115_485} and the He-McKellar-Wilkens phase. Specifically, the He-McKellar-Wilkens phase arises in this geometry from an addition of the Aharonov-Bohm phase acquired for the negatively and positively charged parts of the dipole, i.e. the He-McKellar-Wilkens phase is proportional to the magnetic flux passing through the dipole as it moves around the ring. 

Using this interpretation, we see that the interaction Hamiltonian of the electric dipole of magnitude $d = |{\bf d}|$ in a magnetic field whose vector potential is parallel to the direction of the dipole may be written as the difference in the magnetic interaction of the charges at its ends, \textit{i.e.} as 
\begin{equation}
H_{\rm{int}} = -\frac{\hbar d}{im} {\bf \tilde{A}} \cdot {\bm \nabla} \label{dint}.
\end{equation}
The effective vector potential ${\bf \tilde{A}}$ describes the magnetic interaction of  a dipole in the geometry depicted in Fig. \ref{rydberg}(b) given by
\beq
{\bf \tilde{A}}  = 2\frac{\partial {\bf A}}{\partial r} = B_z \left(-\sin \theta,\cos \theta,0\right),
\eeq 
with ${\bf A} = \frac{1}{2} {\bf{B}}\times{\bf{r}}$ the magnetic vector potential at position ${\bf r}$, which we describe in cylindrical polar co-ordinates $(z, r, \theta)$. $B_z$ is the strength of the $z$-oriented magnetic field, $r$ is the radius in the plane of the toroid, and $\theta$ is the azimuthal angle in the plane of the toroid.
To demonstrate the emergence of a ground state persistent current for the geometry shown in Fig 1(b) below we calculate the energy difference between two states, specifically a state with a $2\pi n$ phase winding in the superfluid phase and a state with $2\pi (n+1)$ phase winding, with integer $n$. For simplicity we assume the the condensate wavefunction has the following property: $|\psi_n({\bf r})|^2=|\psi_{n+1} ({\bf r})|^2$. For this case, and using the interaction Hamiltonian of Eq. (\ref{dint}),  the energy difference between the $n$ and $n+1$ states is given by
\begin{eqnarray}
\Delta E_{n, n+1}&=&\langle \psi_n |H |\psi_n \rangle -\langle \psi_{n+1} |H |\psi_{n+1} \rangle \nonumber \\
&=&-\frac{\hbar^2}{2m}\int d{\bf r} \left[ \psi^{\star}_n \nabla^2 \psi_n - \psi^{\star}_{n+1} \nabla^2 \psi_{n+1}\right] \nonumber \\
&-& \frac{\hbar d}{i m}\int d{\bf r} \left[ \psi^{\star}_n {\bf \tilde{A}} \cdot {\bm \nabla} \psi_n - \psi^{\star}_{n+1} {\bf \tilde{A}} \cdot {\bm \nabla} \psi_{n+1}\right], \nonumber \\
\end{eqnarray}
where terms associated with the trapping potential and interactions have cancelled, since they do not depend on the phase of the wavefunction.

Making an ansatz for the form of the superfluid wavefunction of the following form $\psi_n(x,y,z)=\sqrt{n_0(x,y,z)}\exp[in\arctan(y/x)]$ the energy difference between the $n$ and $n+1$ states simplifies to 
\begin{eqnarray}
\Delta E_{n, n+1}=-\frac{\hbar^2}{2m}(2n+1)\int d{\bf r} \frac{n_0}{r^2} + \frac{\hbar d B_z}{2m}\int d{\bf r} \frac{n_0}{r}.
\label{DE_1}
\end{eqnarray}
To simplify the above we make an ansatz for the functional form of the superfluid density, $n_0$, of the following form
\begin{eqnarray}
n_0(x,y,z) &=&{\tilde n}_0 n_r(r)n_z(z) \label{n0_1}, \\
 n_r(r)  & = &\exp[-(r- r_0)^2/l_r^2] \label{n0_2}\\
n_z(z)&=&\exp[-z^2/l_z^2] \label{n0_3}.
\end{eqnarray}
where $n_0$ is independent of the angle  ($\theta$) in the plane  of the toroid, $n_z(z)$ is associated with an out of plane gaussian density profile of width $l_z$,  $n_r(r)$ is radial density profile and ${\tilde n}_0$ is an normalization constant such that $\int d{\bf r}\, n_0=N$, where $N$ is the total number of atoms in the condensate. The radial density profile, $n_r(r)$, has the form of a toroid, i.e. the peak density is at $r=r_0$ and has a width of $l_r$.  Using Eqs.~(\ref{DE_1},\ref{n0_1},\ref{n0_2},\ref{n0_3}) we find that the energy difference between the $n$ and $n+1$ states simplifies to
\begin{eqnarray}
\Delta E_{n, n+1}&=&\frac{\hbar \pi {\tilde n}_0 l_z}{2m}\Bigg[\pi dB_z l_r \left(1+{\rm Erf}[{\tilde r}_0]\right) \Bigg.\nonumber \\ 
&-&\Bigg.\hbar\sqrt{\pi}(2n+1)\int_0^{\infty} \frac{e^{-({\tilde r}-{\tilde r}_0)^2}}{{\tilde r}} d{\tilde r}\Bigg] \label{DE_2}\\
&\approx&\frac{\hbar \pi {\tilde n}_0 l_z}{2m}\Bigg[\pi dB_z l_r \left(1+{\rm Erf}[{\tilde r}_0]\right) \Bigg.\nonumber \\ 
&+&\Bigg.\frac{\hbar\sqrt{\pi}(2n+1)}{2 {\tilde r}_0^2}\left(e^{-{\tilde r}_0^2}-{\tilde r}_0 \sqrt{\pi}\left(1+{\rm Erf}[{\tilde r}_0]\right)\right)\Bigg], \nonumber \\
\label{DE_approx}
\end{eqnarray}
where ${\rm Erf}[{\tilde r}_0]=(2/\sqrt{\pi})\int_0^{{\tilde r}_0}\exp[-t^2]dt$ is the error function, ${\tilde r} = r/l_r$ and ${\tilde r}_0 = r_0/l_r$. In going from Eq.~(\ref{DE_2}) to (\ref{DE_approx}) we have considered the limit ${\tilde r}_0 \gg 1$, where  
\begin{eqnarray}
\int_0^{\infty} \frac{e^{-({\tilde r}-{\tilde r}_0)^2}}{{\tilde r}} d{\tilde r} \approx -\frac{1}{2 {\tilde r}_0^2}\left(e^{-{\tilde r}_0^2}-{\tilde r}_0 \sqrt{\pi}\left(1+{\rm Erf}[{\tilde r}_0]\right)\right). \nonumber \\
\end{eqnarray}
Numerically, we find that the above approximation is valid for ${\tilde r}_0 > 4$ (note typical experimental values \cite{arxiv:1512.05079} of ${\tilde r}_0$ are  $>5$ ), i.e. the radius of the toroid, $r_0$, is at least four times larger than the width of the condensate in the toroid, $l_r$. In this limit $1+{\rm Erf}[{\tilde r}_0] \approx 2$ and $\exp[-{\tilde r}_0^2] \ll 1$ and hence Eq.~(\ref{DE_approx})  can be simplified: 
\begin{eqnarray}
\Delta E_{n, n+1}&\approx&\frac{\hbar \pi^2 {\tilde n}_0 l_z l_r}{2m}\left[-\frac{\hbar(2n+1)}{r_0} + 2 dB_z \right].
\end{eqnarray}
The point where it is energetically favourable to have a $2\pi (n+1)$ phase winding as apposed to a $2\pi n$ phase winding is defined by $\Delta E_{n, n+1}$ changing sign, from negative to positive. Hence, the magnetic field above which it becomes energetically preferable for a state with $2\pi (n+1)$ phase winding compared to $2\pi n$phase winding is
\begin{eqnarray}
B_z > \frac{\hbar (2n+1)}{2r_0 d}.
\label{eq:bcrit}
\end{eqnarray}
Consider the case of $n=0$; then the critical magnetic field above which the He-McKellar-Wilkens phase induces a persistent ground state flow is given by the simple condition $B_z > \hbar/(2r_0 d)$. If we consider some realistic parameters  $r_0=150 \mu$m \cite{arxiv:1512.05079}, $d=4.2 \times 10^{-30}$Cm, for $^{133}$Cs$^{87}$Rb \cite{PRL113_255301} then $B_z > 0.08\,\text{T} \equiv B_c$, \emph{i.e.} we expect a persistent flow to occur for $B_z>B_c$.

A related experiment was proposed by Sato and Packard \cite{JPC_1503} using a torus of superfluid helium and a similar geometry as that depicted in Fig. \ref{rydberg}(b). The superfluid sample is subjected to a radial electric field generated by charged concentric cylinders inside and around the torus and an axial magnetic field. A large electric field strength of $\sim25\,$kV/cm is required to induce a small dipole moment in the helium atoms, and a large magnetic field (7\,T) is required to observe an appreciable He-McKellar-Wilkens phase. Ultimately these requirements stem from the weak induced dipole moment of the helium atom ($1.5\times 10^{-5}\,$D for a 25\,kV/cm electric field). In our case the large permanent dipole moment of the polar molecule results in a considerable reduction in the required electric and magnetic field strengths.   

Evaluating Eq. (\ref{eq:hmw_def}) for ${\bf B} = B_z {\bf \hat{z}}$, ${\bf d} = d {\bf \hat{r}}$ and $d{\bf l} = r_0 d \theta \boldsymbol{\hat{\theta}}$, it is straightforward to show that for $\phi_\text{HMW} = 2\pi$, $B_z = \frac{\hbar}{r_0 d}$. The factor of two difference between this and the magnetic field strength in Eq. (\ref{eq:bcrit}) arises from the fact that $\phi_\text{HMW}$ need only exceed $\pi$ before the superfluid state with persistent flow becomes energetically favourable.

It is possible to consider the Maxwell dual of the He-McKellar-Wilkens geometric phase, i.e. the Aharonov-Casher effect, for a BEC of magnetic dipoles in a toroidal trap. The geometry to consider is the Maxwell dual of Fig. 1(a), where the electric dipoles ${\bf d}$ are replaced with magnetic dipoles $\boldsymbol{\mu}$, the line of monopoles is replaced with a line of charges and a magnetic field oriented parallel to the line of charges aligns the magnetic dipoles vertically, replacing the electric field.
If a single magnetic dipole is taken once around the line charge, it acquires the AC phase
\begin{equation}
\phi_\text{AC} =  \frac{1}{\hbar c^2}\int \boldsymbol{\mu}\times {\bf E} \cdot d {\bf r},
\end{equation} 
with $c$ the speed of light.

Performing the identical calculation as above, the critical electric field above which it becomes energetically preferable for a $2\pi (n+1)$ phase winding as compared to $2\pi n$ phase winding is
\begin{eqnarray}
E > \frac{\hbar c^2 (2n+1)}{2r_0 \mu},
\end{eqnarray}
where $\mu = |\boldsymbol{\mu}|$ quantifies the strength of the magnetic dipole moment.
For the case of $n=0$ then the critical electric field above which the Aharonov-Casher phase induces a persistent ground state flow is given by the condition $E > \hbar c^2/(2r_0 \mu)$. If we consider  $r_0=150 \mu$m \cite{arxiv:1512.05079}, $\mu=10\,\mu_B$ ($\mu_B$ the Bohr magneton), for $^{164}$Dy \cite{PRL107_190401} then $E > 2\times 10^9 $V/m.

The extreme electric field strength required to see the effects of the Aharonov-Casher phase in magnetic dipolar condensates preclude experimental investigation. However, the induced persistent flow resulting from the He-McKellar-Wilkens effect warrants closer examination. In the case of the Aharonov-Casher effect the electric field strength determines the resulting phase. In order to see an appreciable He-McKellar-Wilkens phase, one requires an electric field of sufficient strength to align the electric dipoles, and the magnetic field strength determines the accumulated phase. Due to the significant dipole moment of polar molecules in the condensate, the requisite electric and magnetic field strengths are substantially less compared to proposals involving helium atoms \cite{JPC_1503}.

However, without a sufficiently strong electric field the `lab accessible' dipole moment $d_\text{lab}$ is somewhat less than the permanent dipole moment $d$. Typically several hundred V/cm \cite{PRL113_255301} up to a few kV/cm \cite{arXiv:0811.4618} is required for  $^{133}$Cs$^{87}$Rb and $^{40}$K$^{87}$Rb respectively to achieve lab-accessible dipole moments of order $d_\text{lab}/d \sim 0.3$ \cite{PRL113_255301}. The requisite geometry of a radially directed field is not trivially realised due to experimental constraints inherent to the ring trap. For instance, it is impractical to place a charged wire \emph{within} the toroidal trap. 

\begin{figure}
	\centering
		\includegraphics[width = \columnwidth]{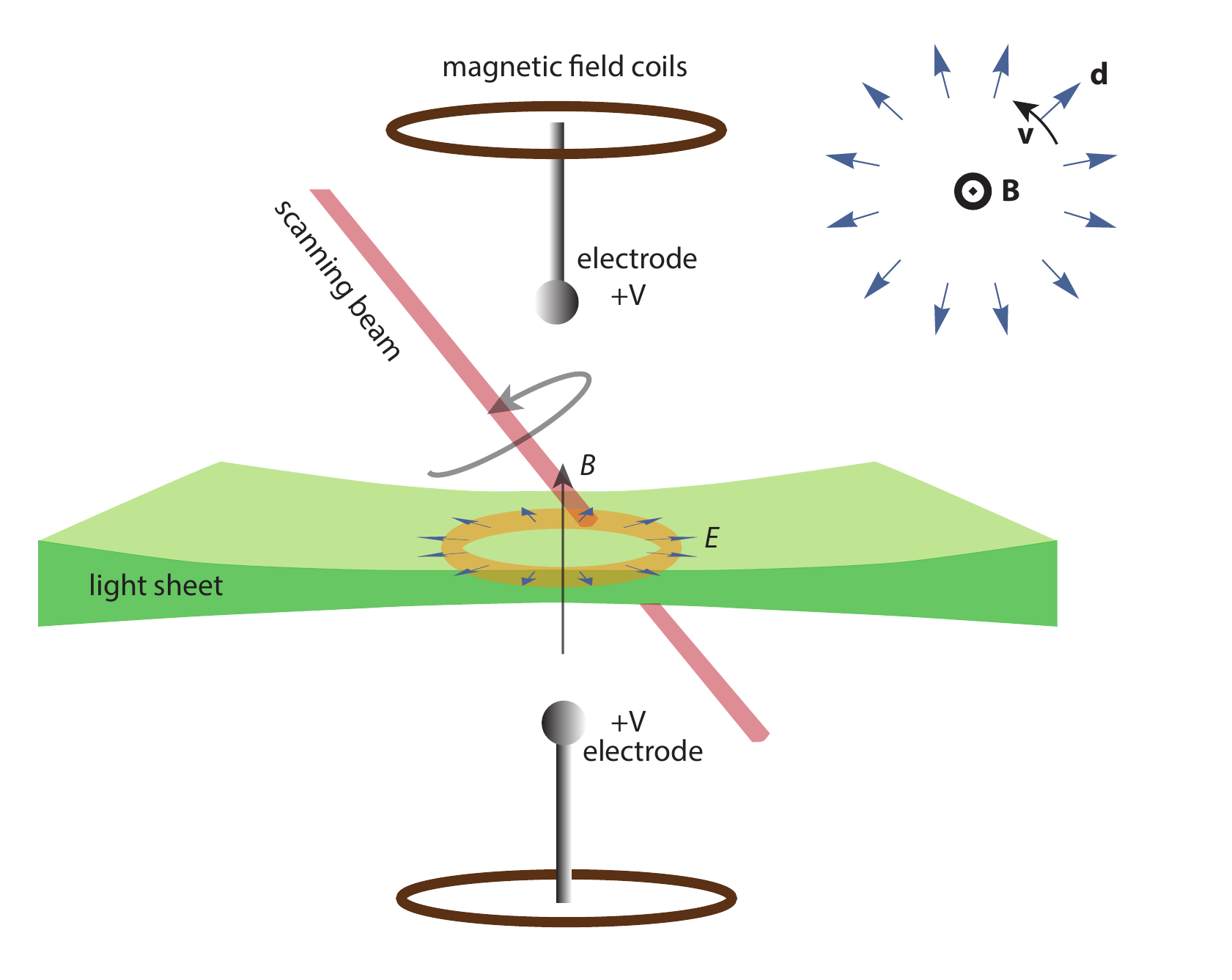}
	\caption{Alternative geometry for observing He-McKellar-Wilkens phase in a BEC of polar molecules. In this example, the ring trap is formed by scanning the position of a beam from an acousto-optic deflector which intersects a light sheet. A pair of charged electrodes above and below the ring trap create a radial electric field of $\sim$kV/cm, polarising the electric dipoles. A pair of current carrying coils located above and below create the required uniform magnetic field along $z$.}
	\label{fig:expgeometry}
\end{figure}

We consider an alternative geometry, where a pair of equally charged electrodes located above and below the ring trap provide a radial electric field at the condensate. For two $50\,\mu$m radius spherical electrodes charged to $200\,$V located $\sim200\,\mu$m above and below the trap, as shown in Fig. \ref{fig:expgeometry}, the electric field is approximately $1.5\,$kV/cm at the condensate. At this field strength, assuming $d_\text{lab}/d\sim0.5$ \cite{PRL113_255301}, the magnetic field strength $B_c$ required to observe a persistent current is increased to $0.16\,$T.

The $z$-oriented magnetic bias field of $B_z>0.16\,$T is achievable with a pair of Helmholtz coils carrying manageably high currents. The ring trap itself can be created using a scanning beam technique \cite{NJP11_043030,arxiv:1512.05079} where condensates in rings up to $300\,\mu$m in diameter have been demonstrated. 

In general, strong electric field gradients exist in our proposed geometry, which give rise to radial forces on the trapped atoms, strong enough to overcome the trapping potential. However, when the electrode separation is set to $2\sqrt{2} r_0$, the electric field gradients vanish.
 
The ideal geometry depicted in Fig. \ref{fig:expgeometry} is still a challenge to implement experimentally. However, a purely radial electric dipole moment is not essential to see an appreciable He-McKellar-Wilkens effect. Any collective departure of the radial electric dipole alignment from the plane of the toroid reduces the cross product in Eq. (\ref{eq:hmw_def}), and can be offset with an increase in the applied magnetic field strength to yield a large enough $\phi_\text{HMW}$. This allows more flexibility in the geometry of electrodes used.    

In conclusion, we have shown that the He-McKellar-Wilkens effect can be used to induce persistent superfluid flow in a BEC composed of electric dipolar molecules. In contrast to the prohibitive experimental requirements to observe a significant Aharonov-Casher phase accumulation, the He-McKellar-Wilkens effect induces a phase shift sufficient to drive superfluid persistent currents for feasible experimental parameters. 
While we have considered the case of condensates trapped in a ring geometry, it is also interesting to consider extending our results to bulk, 2D dipolar condensates trapped in a light sheet. Additionally, in such a geometry, the phase accumulated is now a function of the the radial distance from the center of the trap and is equivalent to considering a non-dipolar BEC in a synthetic magnetic field \cite{NAT462_628} which is increasing radially.

\begin{acknowledgements}
A. A. Wood and A. M. Martin were supported by the Australian Research Council (DP150101704). B.H.J. McKellar was supported in part by the Australian Research Council Grant to the ARC Centre of Excellence for Particle Physics at the Terascale (CoEPP).
\end{acknowledgements}

\end{document}